\documentclass[conference]{IEEEtran}

\usepackage{cite}
\usepackage{amsmath,amssymb,amsfonts}
\usepackage{algorithmic}
\usepackage{graphicx}
\usepackage{textcomp}
\usepackage{xcolor}
\usepackage{algorithm}
\usepackage{listings}
\usepackage{tikz}
\usepackage{hyperref}
\usepackage{amsthm}
\usepackage{cuted}
\ifCLASSINFOpdf
\else
\fi

\begin{document}

\title{PPO-EPO: Energy and Performance Optimization for O-RAN Using Reinforcement Learning \\
\thanks{Telecom Italia sponsors this project.}
}
\author{\IEEEauthorblockN{1\textsuperscript{st} Rawlings Ntassah}
\IEEEauthorblockA{\textit{DISI} \\
\textit{University of Trento}\\
Trento, Italy \\
 \href{mailto:rawlings.ntassah@unitn.it}{rawlings.ntassah@unitn.it}}
\and
\IEEEauthorblockN{2\textsuperscript{nd} Gian Michele Dell'Aera}
\IEEEauthorblockA{\textit{Research and Innovation} \\
\textit{Telecom Italia}\\
Torino, Italy \\
\href{mailto:gianmichele.dellaera@telecomitalia.it}{gianmichele.dellaera@telecomitalia.it}}
\and
\IEEEauthorblockN{3\textsuperscript{rd} Fabrizio Granelli}
\IEEEauthorblockA{\textit{DISI} \\
\textit{University of Trento}\\
Trento, Italy \\
\href{mailto:fabrizio.granelli@unitn.it}{fabrizio.granelli@unitn.it}}

}

\maketitle

\begin{abstract}
Energy consumption in mobile communication networks has become a significant challenge due to its direct impact on Capital Expenditure (CAPEX) and Operational Expenditure (OPEX). The introduction of Open RAN (O-RAN) enables telecommunication providers to leverage network intelligence to optimize energy efficiency while maintaining Quality of Service (QoS). One promising approach involves traffic-aware cell shutdown strategies, where underutilized cells are selectively deactivated without compromising overall network performance. However, achieving this balance requires precise traffic steering mechanisms that account for throughput performance, power efficiency, and network interference constraints.

This work proposes a reinforcement learning (RL) model based on the Proximal Policy Optimization (PPO) algorithm to optimize traffic steering and energy efficiency. The objective is to maximize energy efficiency and performance gains while strategically shutting down underutilized cells. The proposed RL model learns adaptive policies to make optimal shutdown decisions by considering throughput degradation constraints, interference thresholds, and PRB utilization balance. Experimental validation using TeraVM Viavi RIC tester data demonstrates that our method significantly improves the network's energy efficiency and downlink throughput. 
\end{abstract}


% \fi
\IEEEpeerreviewmaketitle

\section{Introduction}
\footnote{This paper has been accepted for publication in the IEEE ICC 2025 Fourth International Workshop on Green and Sustainable Networking. IEEE will publish the final version of this work as: R. Ntassah; G. M. Dell'Aera; F. Granelli "PPO-EPO: Energy and Performance Optimization for O-RAN Using Reinforcement Learning", Proc. IEEE ICC 2025 Fourth International Workshop on Green and Sustainable Networking (GreenNet),  Montreal, Canada, June 2025.} 

\IEEEPARstart{E}{nergy} efficiency is the hallmark of telcos and network operators. Network operators aim to balance reliable service assurance, QoS, and energy sustainability. 

Over half of the total energy usage in telecommunications networks is attributed to the radio access component, with 50\% to 80\% of that energy being consumed by the power amplifier (PA), as noted by the authors in \cite{edler2004energy, chen2014energy, 6157574}. This significant energy demand directly increases operational costs for network operators. For instance, a study in \cite{strinati2010holistic} highlights that Telecom Italia (TIM) is among the highest energy consumers in Italy. These challenges underscore the critical need for a sustainable energy system to enhance profitability and support decarbonization efforts. Disaggregated architectures like O-RAN offer a promising approach by enabling energy-efficient network management and operations through flexible and programmable network components.

The control plane of O-RAN, through the RAN Intelligent Controller (RIC), provides a platform for deploying and managing intelligent models and AI/ML techniques to achieve energy-efficient RAN management. The primary objective is to minimize energy consumption in the RAN while maintaining or enhancing network performance and QoS.

Most energy efficient mechanisms as described in \cite{kundu2024towards, wang2024energy, malik2024achieving, paolini2023energy} concentrate on shutting down idle cells without much concentration on the cells with high interferences and how effective that could contribute to energy reduction network. The handover mechanism for fair UE distribution when a less occupied cell is shut is also not discussed a lot. These are some concerns that require attention since handover could lead to high consumption in the new cell.

To address these challenges, we propose a reinforcement learning (RL) model for traffic steering aimed at improving energy efficiency. Effective utilization of radio resources is crucial in both low and high-traffic scenarios. The RL model reduces resource utilization and operational costs by dynamically handing over UEs from one cell to another and shutting down cells. The proposed model balances energy efficiency and performance gain, adapting to diverse network scenarios and UE distribution patterns. The main contributions are as follows:
\begin{itemize}
    \item Designed an RL model to turn off cells to ensure energy efficiency
    \item Implemented UE handover and redistribution mechanics based on proximity, PRB availability, and interference, ensuring seamless adaptation to cell shutdowns.
    \item Tested model on the dataset generated from TeraVM Viavi RIC tester.
\end{itemize}

\par This article is organized as follows: In Section II we describe the related work, while Section III presents our system model and problem formulation. In Section IV, we discuss the performance and evaluation of our proposed model, and finally, in Section V, our conclusions.

\section{Related work}

Energy consumption and the related costs are the principal reasons for higher CAPEX/OPEX by service providers as described in \cite{strinati2010holistic, edler2004energy}. Holistic efforts to curb energy utilization are vital due to the exponential increase in data transmission and network devices. 

The field of energy efficiency and optimization in wireless communication and Open RAN (O-RAN) systems has gained significant attention in recent years. A detailed overview is provided in \cite{liu2024survey}, highlighting the role of nonconvex optimization, distributed optimization, and learning-based methods in enhancing resource allocation and power control in 5G and 6G networks. While the paper extensively reviews mathematical techniques, its broad focus limits insights into specific real-world applications or implementation challenges.

In \cite{sheikh2024intelligent}, the authors emphasize the integration of RF energy harvesting technologies with advanced signal processing techniques to enhance energy efficiency. This work sheds light on the potential of sustainable power sources like ambient RF energy for next-generation networks but offers limited quantitative validation of its proposals.

The role of virtualization and cloud resources in improving energy efficiency is explored in \cite{demir2024cell}. This study demonstrates the advantages of end-to-end resource orchestration in O-RAN architectures. However, the computational complexity and scalability of the proposed solutions pose challenges for large-scale deployments.

Advances in machine learning-based energy optimization are exemplified in \cite{bordin2024design}. This work introduces deep reinforcement learning (DRL) algorithms for base station management and evaluates them using a realistic simulation framework. Similarly, the authors in \cite{wang2024energy} presented an energy-saving xApp where radio cards (RCs) are switched off to achieve significant energy savings in O-RAN architectures. While both studies highlight the potential of DRL for energy optimization, they primarily rely on simulation data and do not address the challenges of real-world deployment.

In \cite{liang2024enhancing}, the authors introduce xApps designed to optimize power savings by dynamically managing the operational states of radio cards. This work achieves up to 50\% power savings while maintaining service quality. However, the lack of deployment metrics in multi-vendor environments limits its practical insights.

Another approach to energy efficiency is presented in \cite{galeano2024landscape}. This study addresses the cell switch-off problem through multi-objective optimization and demonstrates improved energy savings in dense network environments. However, it does not thoroughly examine the potential impact of such methods on user experience during cell deactivations.

Finally, \cite{kundu2024towards} provides a comprehensive review of state-of-the-art standardization and design strategies for energy-efficient RANs. The article highlights techniques such as full-stack acceleration and network function consolidation. Despite its breadth, the work lacks detailed case studies or quantitative analysis to support the proposed strategies.

\section{System model} \label{env}

\begin{figure}[htbp!]
    \centering
    \includegraphics[width=0.9\linewidth]{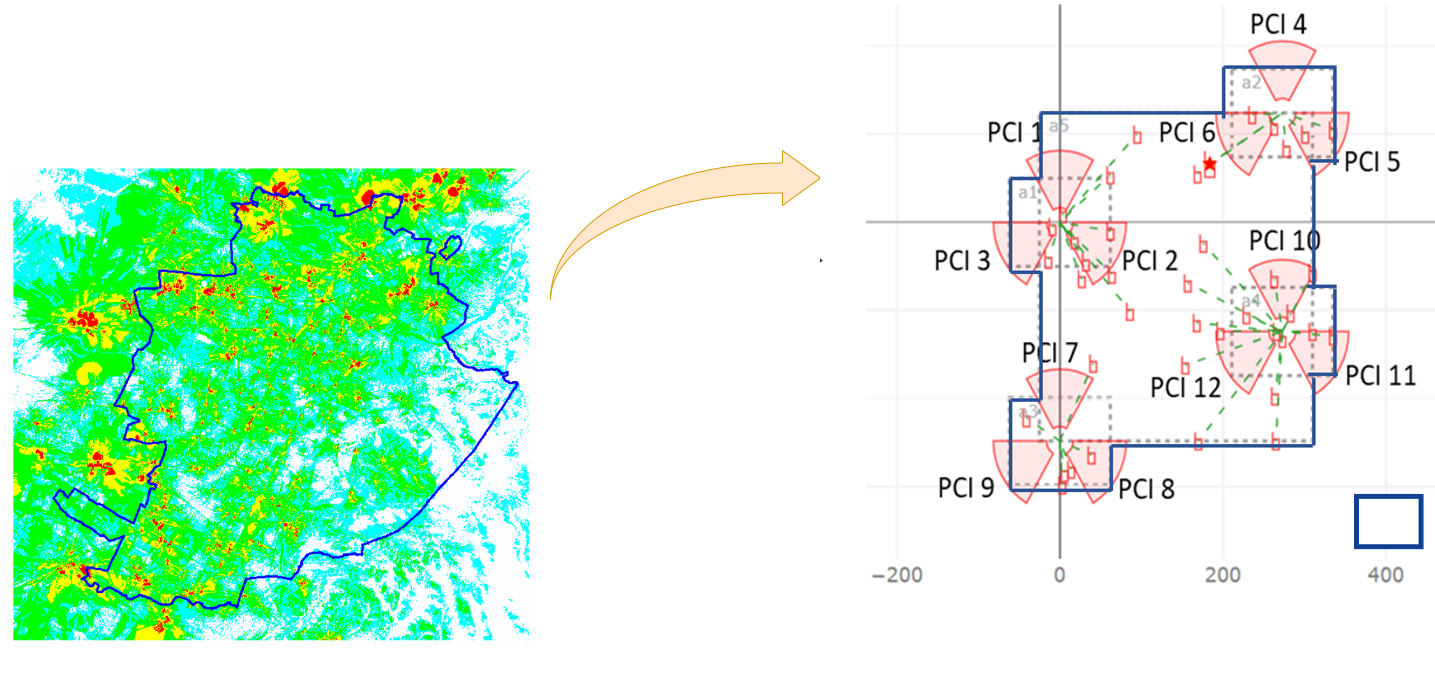}
    \caption{Network system design.}
    \label{networksys}
\end{figure}

The network system consists of one Central Unit (CU) and four Distributed Units (DUs), each connected to the CU. Each DU is associated with three Radio Units (RUs), and the system is designed to represent the network deployment in Turin, Italy, as shown in Fig. \ref{networksys}.

The O-RUs are positioned at a fixed distance from each other. We define an O-RU as $k \in K$ where $K$ is the total number of O-RUs. The network contains a fixed number of user equipment (UEs), represented as \( \mathcal{U} \), which are randomly distributed across the cells.

Each UE is admitted to a cell based on the highest reference signal power ($p_k(u)$), determined using the pathloss equation ($PL_k$) specified by 3GPP for the urban macro (uMa) scenario in \cite{3gppPL}. This is expressed as:
\begin{equation}
    p_k(u) = P_{\text{tx}} - PL_k(d_{k,u}),
\end{equation}
where $P_{\text{tx}}$ is the transmit power of the base station (BS) in dBm, and $PL_k(d_{k,u})$ is the path loss for UE $u$ located at a distance $d_{k,u}$ from BS $k$.

In a given cell \( k \), each UE \( u \) is assigned a downlink (DL) throughput \( R_u \) and a physical resource block (PRB) allocation \( \aleph_u \). The DL throughput received by UE \( u \) is computed as:
\begin{equation}
    R_u = B_u \log_2(1 + \zeta_u),
    \label{eq2}
\end{equation}
where $B_u = \aleph_u B_{\text{prb}}$ represents the effective bandwidth allocated to the UE, \( \aleph_u \) is the number of PRBs assigned to the UE To meet the target throughput \( R_{\text{demand}, u} \), the required number of PRBs for UE \( u \) is calculated as:

\begin{equation}
    \aleph_u = \left\lceil \frac{R_{\text{demand}, u}}{B_{\text{prb}} \log_2(1 + \zeta_u)} \right\rceil,
    \label{prb_demand}
\end{equation}

where
% \begin{itemize}
    % \item 
\( R_{\text{demand}, u} \) is the throughput demand of UE \( u \),
    % \item
   and \( \lceil \cdot \rceil \) is the Ceiling function to ensure the allocation of whole PRBs.
% \end{itemize}
\( B_{\text{prb}} \) is the bandwidth of a single PRB. The signal-to-interference-plus-noise ratio (SINR) of UE \( u \), denoted by \( \zeta_u \), is given by:
\begin{equation}
    \zeta_u = \frac{p_k(u)}{\iota_u + p_{\text{noise}}},
\end{equation}
where \( p_k(u) \) is the power received by UE \( u \), \( \iota_u \) is the interference experienced by the UE, and \( p_{\text{noise}} \) is the noise power at the UE.

The interference \( \iota_u \) for UE \( u \) is expressed as:
\begin{equation}   
    \iota_u = \sum_{i=1}^{k_n} \alpha \left(P_{\text{tx}} - PL_{i}(d_{i,u})\right),
\end{equation}
where \( k_{n} \) is the total number of neighboring cells of cell $k$ and $i$ the neighbor cell. $\alpha$ is a weighting factor based on spatial proximity.

For a cell \( k \) with \( n_k \) UEs, the total and average performance metrics can be summarized as:
\begin{equation}
\label{metric}
\begin{gathered}
\text{Average throughput: } \bar{R}_k = \frac{1}{n_k} \sum_{u=1}^{n_k} R_u, \\
\text{Total throughput: } R_{\text{tot},k} = \sum_{u=1}^{n_k} R_u, \\
\text{Total PRBs: } \aleph_{\text{DL},k} = \sum_{u=1}^{n_k} \aleph_u, \\
\text{Total interference: } I_k = \sum_{u=1}^{n_k} \iota_u.
\end{gathered}
\end{equation}

Here \( n_k \) is the total number of UEs in cell \( k \), where \( n_k \subset \mathcal{U} \), and \( \iota_u \) is the interference power for UE \( u \).

The focus of this work is to shut down cells for energy efficiency which means that the handover (HO) technique is essential. The HO mechanism employed incorporates event A3 as defined by 3GPP \cite{3gppener} for HO execution.  When a cell $k$ is shut down, the total number of UEs ($n_k$) must be redistributed among the neighbor cells. The neighbor cell ($i$) with the highest RSRP ($p_{i}$) becomes a candidate to receive UEs. We compute the interference weight ($\lambda_i$) of the neighbor cell $i$ as $\lambda_i = \frac{1}{||\nu_{i} - \nu_k||^{2} + \epsilon}$. $\nu_{i}$ is the position of neighbor cell $i$, position of the removed cell as $\nu_k$ and constant factor $\epsilon$. A weight is then defined for HO as $\xi = \aleph_{DL, i}. \lambda_i$, where $\aleph_{DL, i}$ is the total PRB used in the neighbor cell. The number of UEs \( n_i \) redistributed to each neighboring cell \( i \) is determined using a multinomial distribution
$n_i \sim \text{Multinomial}(\text{}, \xi)$.

\section*{Problem definition and proposed method}

\subsection{Problem definition}

Each cell consumes energy based on its idle state and active resource utilization in a multi-cell wireless network comprising $K$ cells and \( \mathcal{U} \) UEs. With increasing network densification and traffic variability, energy efficiency has become a critical challenge for modern networks. Shutting down underutilized cells can significantly reduce energy consumption but must be done judiciously to prevent degradation in network performance, including throughput and resource allocation fairness.
% Energy consumption is vital in network management and operations. 

In our scenario, we consider the power consumption of a cell $k$ to consist of the sum of the power in an idle state (without load) plus the transmission power $P_{\text{Tx}, k}$ by $\eta$. The total power consumed ($P_{k}$) by cell $k$ is described in eq. \ref{power}
\begin{equation}
P_k = P_{\text{idle}} + P_{\text{Tx}, k}/\eta
\label{power}
\end{equation}  where $P_{\text{Tx}, k} = \aleph_{\text{DL, k}} \cdot P_{\text{tx}}$,  $\eta$ is the power amplifier efficiency and $P_\text{tx}$ is the maximum transmission power. The energy efficiency ($EE_k$) of the cell $k$ is denoted by 

%%%%%%%%%%%%%%%%%%%%%%%%%%%%%%%%%%%%%%%%%%%%%%%%%%%%%%%%%%%%%%%%%%%%%%%%%%%%%%%%%%%
\begin{equation}
EE_{k} = \frac{\bar{R}_{k}}{P_k}  \implies  \frac{\bar{R}_{k}}{\aleph_{\text{DL, k}} \cdot P_{\text{tx}} \cdot \eta^{-1} + P_{\text{idle}}}
\label{totpow}
\end{equation}

% In this case, the rate $R$ is defined using the bandwidth and the SINR and can be represented as \[ R_u = B.\log_2(1 + \zeta) \] where $\zeta = \frac{p_u}{\iota_u + p_{noise}}$ and $p_u$ is the power received by the UE and defined as $P_{max} - PL$ using $PL$ (path loss) as defined in \cite{3gppPL} for uMa, $\iota_u$ the interference, and $p_{noise}$ is the noise power at the UE. $R_k$ is the sum of throughput $R_u$ as defined in \ref{metric}
%%%%%%%%%%%%%%%%%%%%%%%%%%%%%%%%%%%%%%%%%%%%%%%%%%%%%%%%%%%%%%%%%%%%%%%%%%%

Hence, to effectively manage the energy and performance of our system, it is essential to analyze both the performance gain ($g_{\text{k}}$) and the power gain ($P_{\text{gain, k}}$) of the system. In the computation of the  $G_{\text{perf, k}}$, we deduce the difference between the average DL throughput ($\bar{R}_{\text{after}}$) of the network system after a cell $k$ is shut off and the average throughput ($\bar{R}_{\text{before}}$) shutting down that particular cell and divide the difference by $\bar{R}_{\text{before}}$. 

Similarly, we compute the power gain $P_{\text{gain, k}}$ by taking the difference between the average power consumed by the system  ($P_{\text{avrg, before}}$) and the power after ($P_{\text{avrg, after}}$) divided by the maximum transmission power ($P_{\text{max}}$) as shown in eq. \ref{energyEF}. 

\begin{equation}
g_{\text{k}} = \frac{\bar{R}_{\text{k, after}} - \bar{R}_{\text{k, before}}}{\bar{R}_{\text{k, before}} + \alpha}, \quad G_{\text{perf}} = \frac{1}{K}\sum_{k=1}^{K} g_{\text{k}}
\label{gain}
\end{equation}

% \begin{equation}
% G_{\text{perf, k}} = \frac{\bar{R}_{\text{after}} - \bar{R}_{\text{before}}}{\bar{R}_{\text{before}}}  %, \quad %G_{\text{prb\_gain, k}} = \frac{\bar{\aleph}_{\text{before}} - \bar{\aleph}_{\text{after}}}{\bar{\aleph}_{\text{before}}}.
% \label{gain}
% \end{equation}

The average power consumed $P_{avrg}$ and average throughput $R_{total}$ in the entire network are given by 
\begin{equation}
\begin{gathered}
P_{avrg} = \frac{1}{K}\sum_{k=1}^{K} P_{k} \quad
R_{avrg} = \frac{1}{K}\sum_{k=1}^{K}  R_{total, k}
\label{tpow}
\end{gathered}
\end{equation}.
%where $\aleph_{\text{DL, used}} = \sum_{k=1}^{K} \aleph_{\text{used, k}}$ and $\aleph_{\text{DL, used}} \subseteq  $

% \begin{equation}
% R_{avrg} = \frac{1}{K}\sum_{k=1}^{K}  R_{total, k} 
% \label{thrp}
% \end{equation}

% \begin{equation}
% P_{total} = \sum_{k=1}^{K} P_{k} 
% \label{totpow}
% \end{equation}

The energy efficiency of the network can then be defined as illustrated in eq. \ref{enereff}
\begin{equation}
EE_{total} = \frac{R_{avrg}}{P_{avrg}}
\label{enereff}
\end{equation} where $R_{avrg}$ is the throughput of the system.

In eq. \ref{energyEF}, the power gain when cell $k$ is switched off is defined. 

\begin{equation}
% \begin{gathered}
P_{gain} = \frac{P_{avrg, before}-{P_{avrg, after}}}{P_{max}}  
% \\ 
% \implies \frac{\frac{1}{K} \sum_{k=1}^{K} P_{\text{idle, k}} + \frac{\aleph_{\text{total, used, before}}}{K \cdot \eta} \cdot P_{\text{max}} - \frac{1}{K-1} \sum_{k \neq k'} P_{\text{idle, k}} + \frac{\aleph_{\text{total, used, after}}}{(K-1) \cdot \eta} \cdot P_{\text{max}}}{\frac{1}{K} \sum_{k=1}^{K} P_{\text{idle, k}} + \frac{\aleph_{\text{total, used, before}}}{K \cdot \eta} \cdot P_{\text{max}}}
\label{energyEF}
% \end{gathered}
\end{equation}
% \begin{strip}
% \begin{equation*}
% \begin{gathered}
% P_{gain} = \frac{P_{avrg, before}-{P_{avrg, after}}}{P_{max}} \implies \frac{\frac{1}{K} \sum_{k=1}^{K} P_{\text{idle, k}} + \frac{\aleph_{\text{total, used, before}}}{K \cdot \eta} \cdot P_{\text{max}} - \frac{1}{K-1} \sum_{k \neq k'} P_{\text{idle, k}} + \frac{\aleph_{\text{total, used, after}}}{(K-1) \cdot \eta} \cdot P_{\text{max}}}{\frac{1}{K} \sum_{k=1}^{K} P_{\text{idle, k}} + \frac{\aleph_{\text{total, used, before}}}{K \cdot \eta} \cdot P_{\text{max}}}
% \label{energyEF}
% \end{gathered}
% \end{equation*}
% \end{strip}

where:
\[
P_{\text{avrg, before}} = \frac{1}{K} \sum_{k=1}^{K} (P_{\text{idle, k}} + \frac{\aleph_{\text{DL, k, before}}}{\eta} \cdot P_{\text{tx}})\], \text{and}
\[\
P_{\text{avrg, after}} = \frac{1}{K-1} \sum_{k \neq k'} (P_{\text{idle, k}} + \frac{\aleph_{\text{DL, k, after}}}{ \eta} \cdot P_{\text{tx}}).
\]

% Hence, the energy efficiency gain $E_{gain}$ of the system when cell $k$ is switched off is given as follows: 

% \begin{equation}
% EE_{gain} = \frac{G_{\text{perf, k}}}{P_{gain, k}}
% \label{eneEE}
% \end{equation}

This becomes an optimization problem where we must find the cell ($k$) to switch off to reduce energy consumption while maximizing the performance gain. The objective function is therefore defined as

% \begin{equation}
% \max_{x_k} \quad \sum_{k=1}^{K} x_k \left(\omega_{perf} \cdot G_{\text{perf}} + \omega_{power} \cdot P_{\text{gain, k}} \right),
% \label{object}
% \end{equation}
% where $x=1$ indicates cell $k$ is shut down, $x_k \in {0, 1}$ 
% % \begin{equation}
% % \max_{x_k} \quad \sum_{k=1}^{K} x_k \left(w_{\text{EE}} G_{\text{perf}} + w_{\text{Perf, k}} G_{\text{energyE, k}} \right),
% % \label{object}
% % \end{equation}
% subject to:
% \[
% \bar{\aleph}_{\text{avrg, before}} > \bar{\aleph}_{\text{avrg, after}}, \quad \forall k,
% \]
% \[
%     \delta \cdot \bar{R}_{\text{before}} \leq \bar{R}_{\text{after}}, \quad \forall k,
% \]
% \[
% x_k \in \{0, 1\}, \quad \forall k.
% \]

\begin{equation}
\max_{x_k} \quad \sum_{k=1}^K x_k \left( \omega_{\text{perf}} \cdot G_{\text{perf, k}} + \omega_{\text{power}} \cdot P_{\text{gain, k}} \right),
\label{optimize}
\end{equation}
st:
\[
\bar{R}_{\text{avrg, after}} \geq \delta \cdot \bar{R}_{\text{avrg, before}}, \quad \forall k,
\]
\[
\aleph_{\text{avrg, after}} \leq \aleph_{\text{avrg, before}}, \quad \forall k,
\]
% \[
% \left| \aleph_{\text{used, i}} - \frac{\aleph_{\text{total, used}}}{K-1} \right| \leq \epsilon, \quad \forall i \neq k,
% \]
% \[
% \frac{\bar{R}_{\text{max}}}{\bar{R}_{\text{min}}} \leq \gamma, \quad \forall k,
% \]
\[
I_{\text{after}} \leq I_{\text{threshold}}, \quad \forall k,
\]
% \[
% EE_{\text{after}} \geq \beta \cdot EE_{\text{before}}.
% \]

where $\delta \in [0,1]$ and $\aleph_{avg} = \frac{1}{K}\sum_{k=1}^{K} \aleph_{\text{DL, k}}$ 
%and $\beta \in [0,1]$ as an efficiency threshold.

% where \( w_{\text{EE}}, w_{\text{Perf}} \) are the weights for energy efficiency and performance gain, respectively.

\section{Performance and evaluation}

The network architecture features a single Central Unit (CU) that connects to four Distributed Units (DUs). Each DU links to three Radio Units (RUs). This structure mirrors the network deployment implemented in Turin, Italy, as illustrated in Fig. \ref{networksys}.
We consider a network environment of 12 RUs in a $400 \times 400  m^{2}$ area. The network is setup as described in Section \ref{env}. The parameters used in building the environment are presented in Table \ref{table2}. The training of our model was done using the stable baseline PPO \cite{raffin2021stable} model with an HPC cluster of a single node with 2 GPUs, 4 CPUs per task, and 15GB memory.

\begin{table}[htbp!]
    % \centering
    \caption{Table of network parameters}
    \resizebox{\columnwidth}{!}{%
    \begin{tabular}{|p{5cm}|c|} \hline
       \textbf{Parameter}  &  \textbf{value} \\ \hline
        Cell radius (m) &	250  \\  \hline
        Inter-cell distance (m) & 500 \\ \hline
        Number of neighbors &	4  \\ \hline
       Minimum PRB  	& 10 \\ \hline
       Maximum PRB & 100 \\ \hline  
        Number of UEs	& 40  \\ \hline
        Minimum throughput (Gb)	&  0.01   \\   \hline       
        Maximum throughput (Gb)	& 0.1 \\ \hline
        Learning rate &	\( 1 \times 10^{-5} \)  \\  \hline
        Batch size & 64 \\ \hline
         Discount factor (\( \gamma \)) & 0.99 \\ \hline
        GAE (\( \lambda \))	& 0.95 \\ \hline
    \end{tabular} 
    }
    \label{table2}
\end{table}

\subsection{Model training}

% To train the PPO model, the parameters listed in Table \ref{table3} are adopted.

% \begin{table}[htbp!]
%     \centering
%     \caption{PPO model parameters}
%     \resizebox{\columnwidth}{!}{%
%     \begin{tabular}{|p{10cm}|c|} \hline
%        \textbf{Parameter}  &  \textbf{value} \\ \hline
%         Learning rate &	\( 1 \times 10^{-5} \)  \\  \hline
%         Batch size & 64 \\ \hline
%         Discount factor (\( \gamma \)) & 0.99 \\ \hline
%        GAE (\( \lambda \))	& 0.95 \\ \hline
      
%     \end{tabular} }
%     \label{table3}
% \end{table}

% \begin{itemize}
%     \item \textbf{Learning rate:} \( 1 \times 10^{-6} \) (controls the step size for updating model parameters).
%     \item \textbf{Number of steps:} 2048 (determines how many steps the agent collects before updating the policy).
%     \item \textbf{Batch size:} 64 (number of samples used for each gradient update. A batch size of 32 is also efficient).
%     \item \textbf{Discount factor (\( \gamma \)):} 0.99 (balances the importance of future rewards compared to immediate rewards).
%     \item \textbf{GAE (\( \lambda \)):} 0.95 (balances bias and variance in advantage estimation).
%     \item \textbf{Number of training timestamps (Millions):} 5 or 10 (depending on the configuration).
%     \item \textbf{Number of epochs:} 100 (number of times the policy network iterates over the collected data for each update).
%     \item \textbf{Policy type:} Multi-Layer Perceptron Policy (MlpPolicy) 
% \end{itemize}

At the beginning of the simulation (time step 0), the network environment is set up with a predetermined configuration with fixed network infrastructure, an initial distribution of user equipment (UE), and comprehensive metric calculations for each UE and cell.
The network topology remains static throughout the training process. However, there is a reconfiguration of the network at the start of each subsequent episode by dynamically redistributing user equipment across different cells. This ensures that UEs are distributed differently adding dynamism and network scenarios for the agent to learn.

\begin{figure}[htbp!]
    \centering
    \includegraphics[width=9cm]{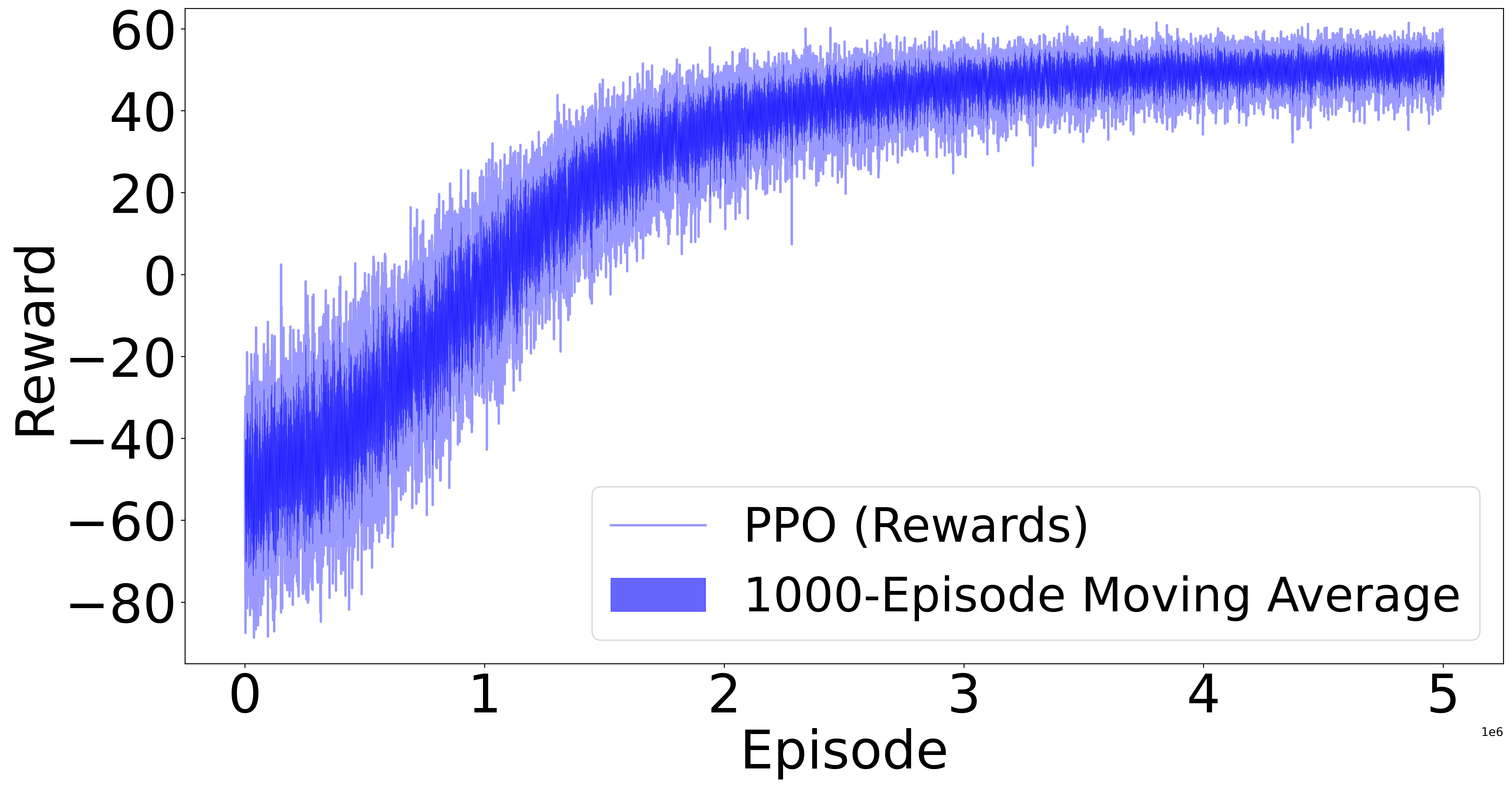}
    \caption{PPO training rewards for 5 million episodes}
    \label{pporew}
\end{figure}

The agent analyzes the metrics associated with network cells and user equipment to strategically determine which cell to deactivate. After making this decision, the agent calculates the corresponding rewards based on the outcomes of its actions and subsequently refines its policy. This iterative learning process continues until the algorithm converges on an optimal strategy for network management, as illustrated in Fig. \ref{pporew}. We trained our model over 5 million episodes, and the resulting reward trends demonstrate the gradual and effective improvement in the agent's decision-making policy.

\subsection{Model Validation}
For validation, we applied the proposed PPO-based reinforcement learning model to a dataset generated using the TeraVM Viavi RIC Tester at Telecom Italia. The experimental setup ensured that the environment and data processing pipeline were consistent with the training procedure, maintaining uniform feature selection and preprocessing steps. The processed data was then fed into the trained PPO model, which predicted the optimal cell shutdown decisions across different network scenarios.

\begin{figure}[htbp!]
    \centering
    \includegraphics[width=9cm]{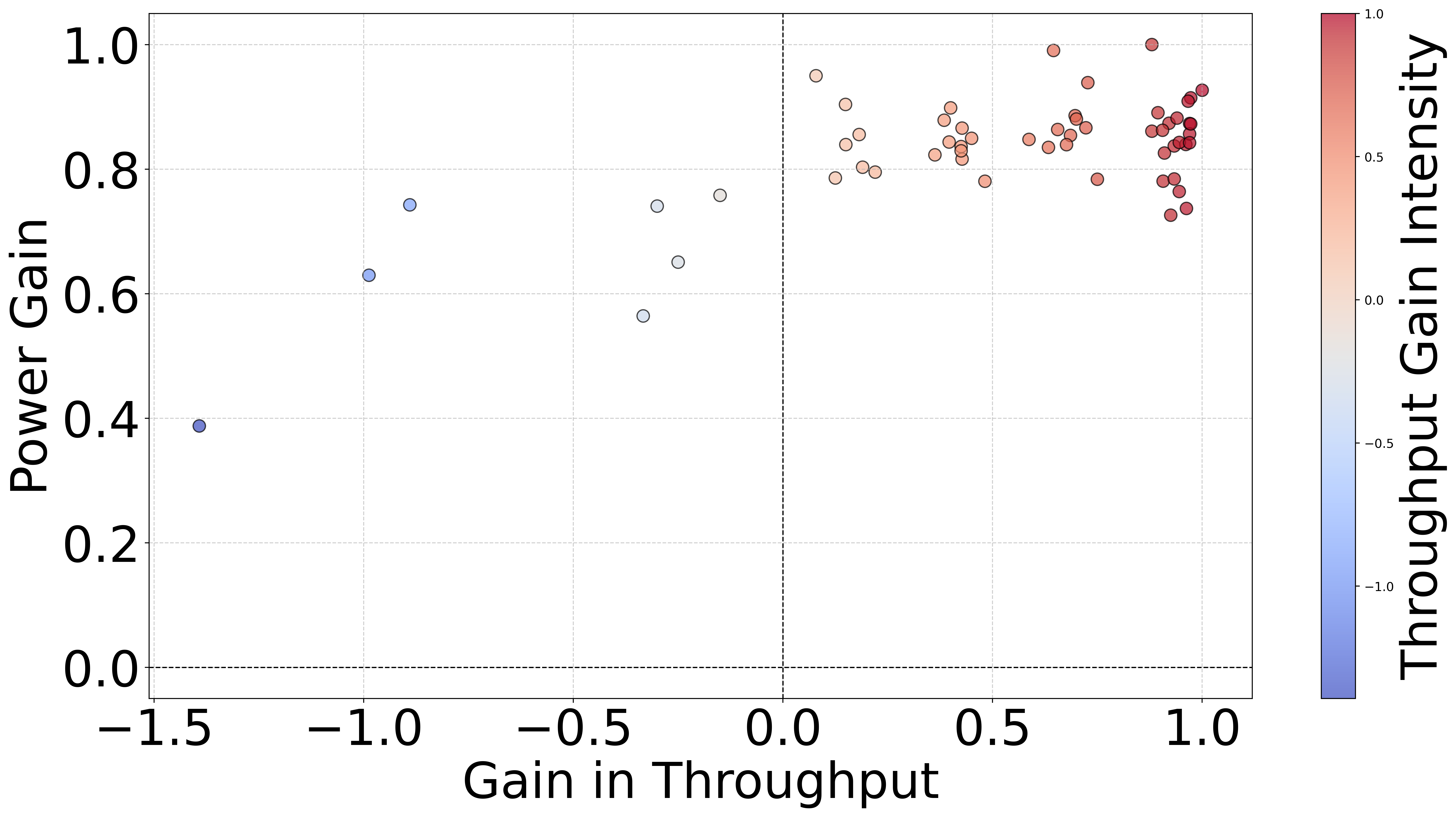}
    \caption{Distribution of both power and throughput gain showing the heatmap}
    \label{perf_pow}
\end{figure}

Following the predicted shutdown actions, we analyzed the relationship between throughput gain and power gain, both of which are key parameters in our optimization objective function defined in \ref{optimize}. The results, illustrated in Fig. \ref{perf_pow}, show the distribution of throughput and power gains, highlighting the effectiveness of our optimization approach in enhancing throughput while minimizing power consumption. However, the presence of negative throughput gains indicates instances where network decisions led to localized performance degradation, which aligns with the allowable throughput degradation constraints in our objective function. This trade-off is influenced by the model’s weight assignments, where energy efficiency was prioritized over throughput performance with $w_{perf} = 0.4$ and $w_{power}= 0.6$. The overall trend underscores the importance of carefully balancing power savings and throughput maximization, reinforcing the need for intelligent network policy optimization to achieve sustainable and high-performance O-RAN deployments.

\begin{figure}[htbp!]
    \centering
    \includegraphics[width=9cm]{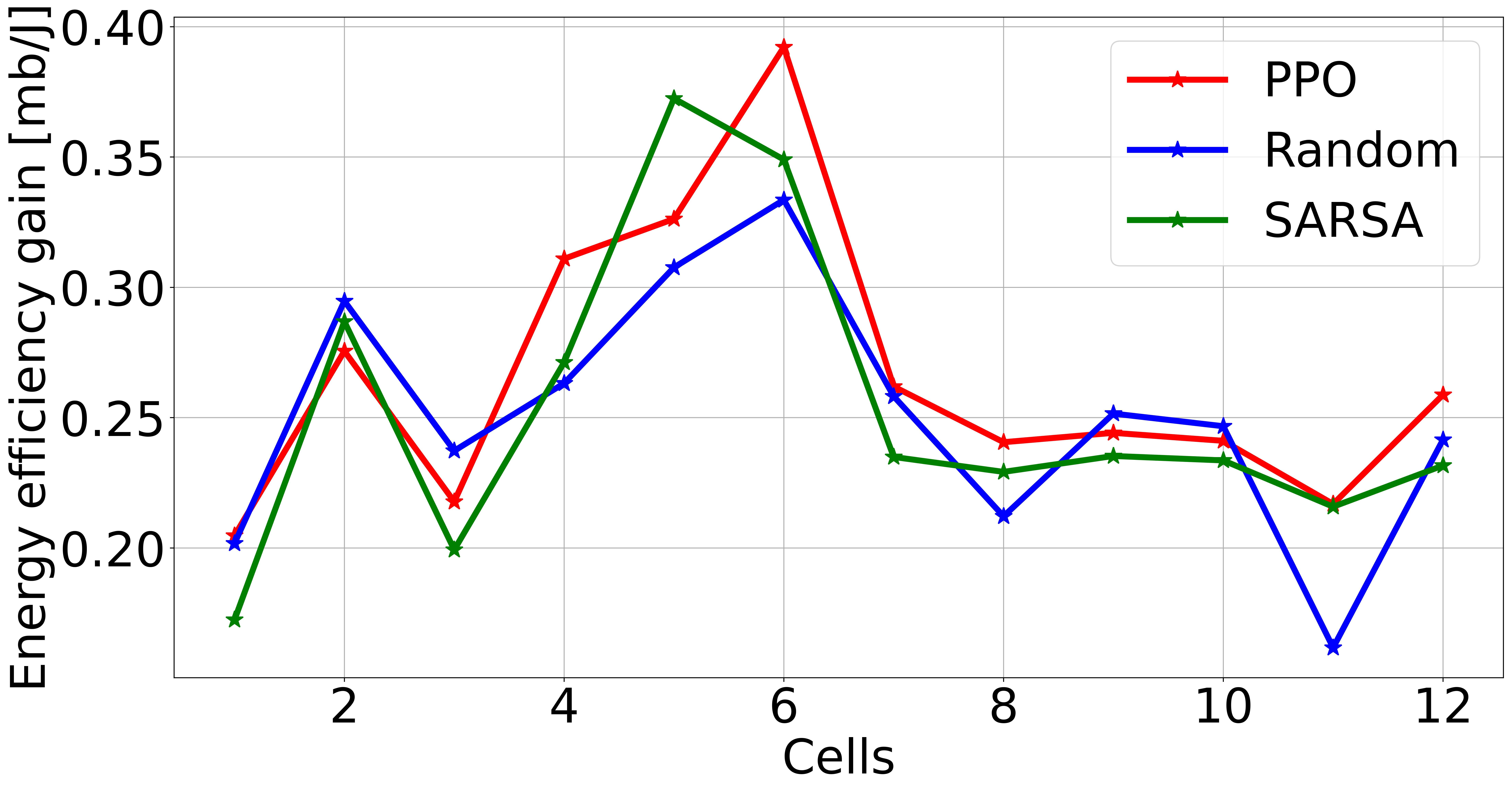}
    \caption{The energy gain (The ratio of the rate and the power for the three methods)}
    \label{energy}
\end{figure}

% as discussed in \cite{ozturk2021energy}
Fig. \ref{energy} evaluates the energy efficiency gains across different cells using three decision-making models. PPO, State–action–reward–state–action (SARSA), and a Random Selection strategy. The results demonstrate that PPO consistently outperforms SARSA and the Random Selection method, indicating its capability to make intelligent cell shutdown and UE reallocation decisions. By dynamically adjusting power and PRB allocations, PPO maximizes throughput while minimizing energy wastage, leading to superior energy efficiency. SARSA, though less effective than PPO, provides more stable performance with fewer fluctuations, making it a viable option where predictability is preferred. In contrast, the Random Selection approach performs the worst, as arbitrary shutdown decisions lead to inefficient power usage and suboptimal UE distributions, reinforcing the necessity of intelligent optimization techniques.

% A notable observation in Fig. \ref{} is the correlation between UE distribution and PRB utilization, where PPO's energy efficiency improvements are linked to optimized resource allocation. However, in certain cells, PPO does not significantly outperform SARSA, suggesting that the model may sometimes prioritize throughput gains over energy efficiency. This highlights the need for fine-tuning PPO’s reward function to ensure a balanced trade-off between power savings and performance. The overall findings validate that machine learning-based optimization strategies, particularly PPO, enhance energy-efficient network operations, while non-intelligent selection strategies, such as Random Selection, result in inefficient and unpredictable outcomes.

\begin{figure}[htbp!]
    \centering
    \includegraphics[width=9cm]{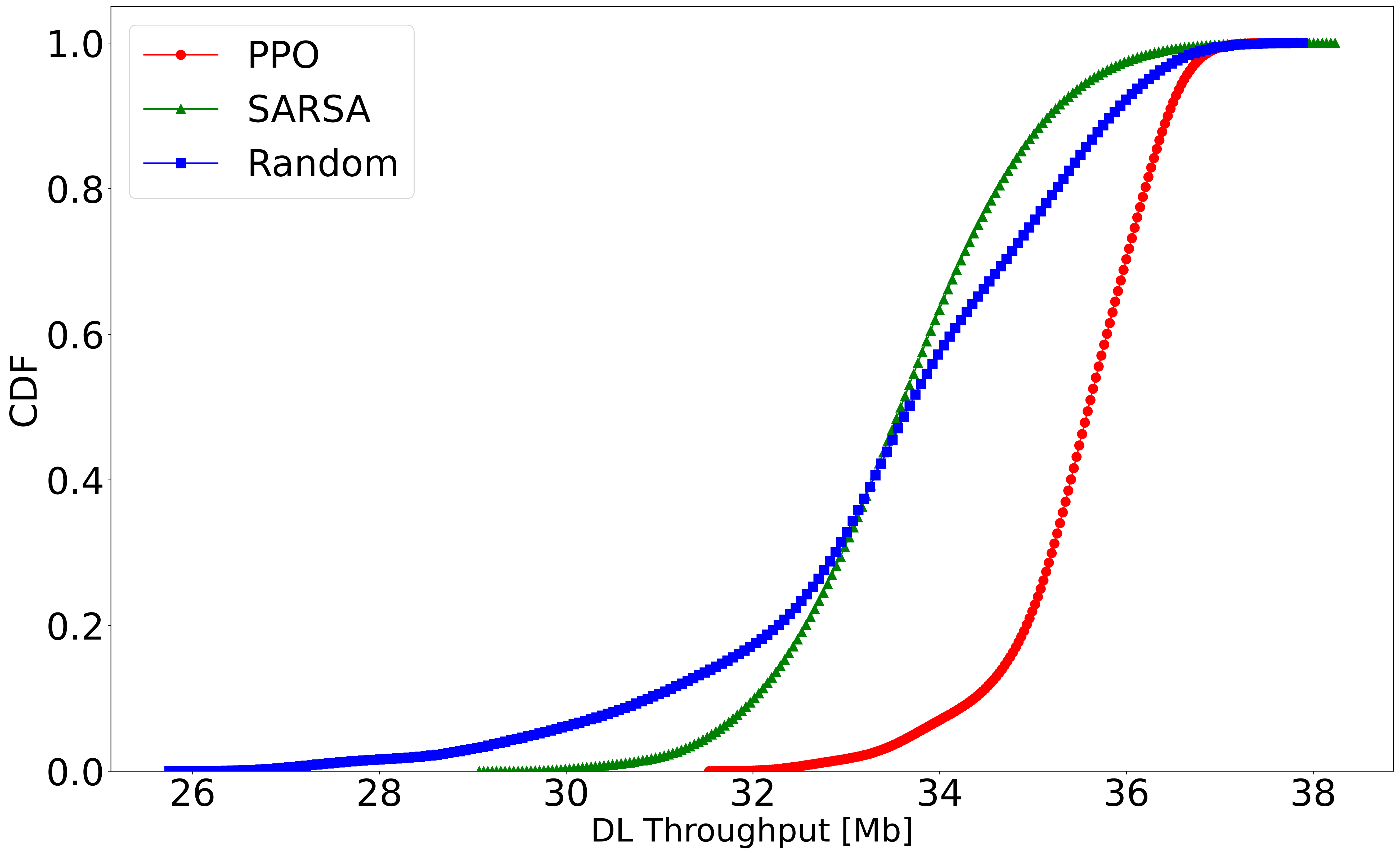}
    \caption{CDF of the throughput of the network for the different methods}
    \label{thrpp}
\end{figure}

Fig. \ref{thrpp} presents the cumulative distribution function (CDF) analysis of downlink throughput. The results demonstrate that PPO significantly outperforms the baseline models by maintaining higher throughput as seen in the CDF analysis. This is because the PPO model has learned the optimal policy to shut down cells to achieve higher throughput and resource distribution, effectively reducing network congestion and enhancing data rates. The PPO enables a larger proportion of UEs to experience improved throughput by efficiently distributing PRBs and optimizing power allocation. While SARSA performs better than Random Selection, its faster CDF rise suggests that it does not allocate throughput as optimally as PPO. The Random Selection model exhibits low performance, with a rapid increase in its CDF curve. Most UEs are concentrated below 32 Mbps, highlighting the inefficiency of random decision-making in network optimization.

% Although SARSA follows a trend similar to PPO, it does not fully maximize throughput due to its more conservative decision-making approach. This results in slightly lower throughput values, suggesting that SARSA, while more stable, lacks the dynamic adaptability required for aggressive performance improvements. In contrast, the Random Selection model exhibits the worst performance, with a rapid increase in its CDF curve, indicating that a significant number of UEs experience poor throughput. This reinforces the necessity of intelligent scheduling and load-balancing techniques, as random decision-making leads to inefficient spectrum utilization. Furthermore, the strong correlation between throughput and energy efficiency highlights the need for joint optimization strategies, where PPO outperforms the other models by balancing power consumption with high transmission rates. These findings suggest that reinforcement learning-based optimization, particularly PPO, is a viable solution for dynamically managing network resources and improving user experience in real-time.

\begin{figure}[htbp!]
    \centering
    \includegraphics[width=9cm]{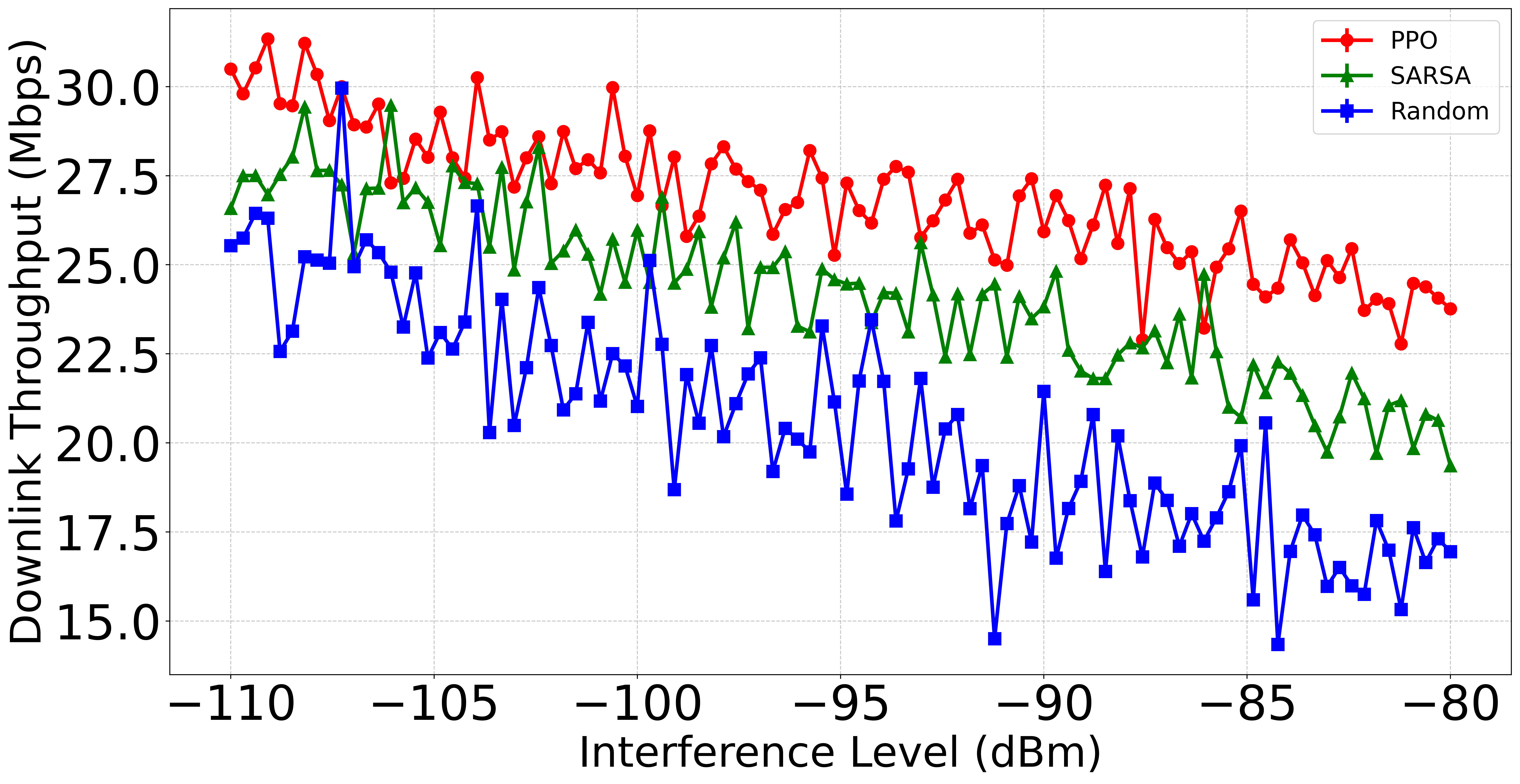}
    \caption{This graph illustrates the relationship between the throughput and interference}
    \label{thrInte}
\end{figure}

The relationship between the interference and throughput highlights the effectiveness of the PPO method in managing network interference while maintaining downlink throughput. The PPO consistently achieves high throughput across all interference levels, demonstrating its ability to make intelligent cell shutdown decisions that minimize performance degradation as shown in Fig. \ref{thrInte}. SARSA follows a similar but slightly lower throughput per interference as compared to the PPO method, indicating that while it improves throughput compared to Random Selection, its on-policy learning limits adaptability, making it more sensitive to interference. Random Selection exhibits the sharpest drop in throughput as interference increases, which explains that arbitrary cell shutdown decisions lead to poor UE distribution, inefficient spectrum usage, and higher performance degradation.

\section{Conclusion}
This study presents a reinforcement learning-based framework for optimizing traffic steering and energy efficiency in mobile networks. The proposed PPO-based approach successfully maximizes energy efficiency and throughput gains while ensuring that critical performance constraints are met. By incorporating PRB allocation constraints, interference thresholds, and throughput degradation limits, the model effectively prevents unnecessary performance deterioration while minimizing energy consumption. 

For future work, we suggest Hybrid reinforcement learning models, which could combine SARSA’s stability with PPO’s adaptability, thereby improving overall network robustness and Hierarchical reinforcement learning (HRL), which could enable more dynamic traffic management, predictive modeling, and adaptive cell shutdown policies. The inclusion of hierarchical RL for dynamic network analysis is particularly promising, as it could facilitate real-time traffic prediction and context-aware policy adjustments, further enhancing O-RAN adaptability. 

% or


\section*{Acknowledgement}
  This work was jointly supported by Telecom Italia S.p.A. under the UniversiTIM framework and the Italian National Inter-University Consortium for Telecommunications (CNIT). The research was partially funded by the European Union's Horizon Europe research and innovation program under the HORSE - Holistic, Omnipresent, Resilient Services for Future 6G Wireless and Computing Ecosystems – G.A. 101096342.

\bibliographystyle{IEEEtran}
%\bibliography{IEEEabrv,references}
\typeout{}\bibliography{reference}



%\bibliographystyle{IEEEtran}

\end{document}